\documentclass[conference]{IEEEtran}
\IEEEoverridecommandlockouts
\usepackage{cite}
\usepackage{amsmath,amssymb,amsfonts}
\usepackage{graphicx}
\usepackage{textcomp}
\usepackage{xcolor}
\usepackage{url}
\usepackage{balance}
\usepackage{booktabs}
\usepackage{siunitx}
\usepackage{multirow} 
\usepackage{balance}

\def\BibTeX{{\rm B\kern-.05em{\sc i\kern-.025em b}\kern-.08em
    T\kern-.1667em\lower.7ex\hbox{E}\kern-.125emX}}

\begin{document}

\title{Scaling 5G-TSN Bridges: Operating Regimes, Scheduling, and Time Synchronisation Under Heterogeneous Industrial Traffic
\thanks{This publication has emanated from research conducted with the financial support of Research Ireland under Grant number 13/RC/2077 P2. For the purpose of Open Access, the author has applied a CC-BY public copyright license to any Author Accepted Manuscript version arising from this submission}
}

\author{
\IEEEauthorblockN{Mohamed Seliem}
\IEEEauthorblockA{\textit{School of CS and IT} \\
\textit{University College Cork}\\
Cork, Ireland \\
MSeliem@ucc.ie}
\and
\IEEEauthorblockN{Utz Roedig}
\IEEEauthorblockA{\textit{School of CS and IT} \\
\textit{University College Cork}\\
Cork, Ireland \\
U.Roedig@ucc.ie}
\and
\IEEEauthorblockN{Cormac Sreenan}
\IEEEauthorblockA{\textit{School of CS and IT} \\
\textit{University College Cork}\\
Cork, Ireland \\
Cormac.Sreenan@ucc.ie}
\and
\IEEEauthorblockN{Dirk Pesch}
\IEEEauthorblockA{\textit{School of CS and IT} \\
\textit{University College Cork}\\
Cork, Ireland \\
Dirk.pesch@ucc.ie}
}

\maketitle

\begin{abstract}
3GPP Release~16 enables a 5G system to operate as a transparent
IEEE~802.1 TSN bridge, but its scalability under heterogeneous
industrial workloads remains insufficiently characterised. This paper
uses the nascTime framework on OMNeT++/Simu5G to evaluate how many TSN
endpoints a single 5G~NR cell can bridge before per-flow QoS degrades.
We model closed-loop control, machine vision, bulk telemetry, and
IEEE~802.1AS traffic over a four-bearer SDAP architecture, varying the
number of endpoints from 1 to~40, MAC scheduler, radio bandwidth
(\SI{10}{\mega\hertz} and \SI{20}{\mega\hertz}), and channel model.
Results show three operating regimes. Below saturation, non-DRR
schedulers perform similarly; near saturation, QoS-aware PF reduces
critical-flow P99 latency by up to two orders of magnitude relative to
channel-aware and fairness-based schedulers; and under overload, only
QoS-PF maintains near-complete delivery for the highest-priority
traffic. Across the two evaluated bandwidths, the saturation threshold
approximately doubles when bandwidth doubles. We also show that
isolating IEEE~802.1AS/gPTP traffic on a dedicated high-priority radio
bearer reduces clock-servo instability, although endpoints carrying
lower-priority data still experience elevated synchronisation delay
under saturation because of reduced MAC scheduling frequency. Finally,
the evaluated sub-6\,GHz, \SI{30}{\kilo\hertz}-SCS configuration
exhibits an effective latency floor of approximately
\SI{2.25}{\milli\second}, indicating that sub-\SI{3}{\milli\second}
TSN deadlines may require radio-configuration changes such as
configured grants or higher numerology.
\end{abstract}


\section{Introduction}\label{sec:intro}

3GPP Release~16 specifies how a 5G system can operate as a
transparent IEEE~802.1 Time-Sensitive Networking (TSN) bridge,
mapping TSN priority classes through 5G QoS flows and radio bearers
while supporting IEEE~802.1AS transparent-clock behaviour~\cite{3gpp23501}.
Although early testbeds demonstrate feasibility at small scale, a key
deployment question remains unresolved: \emph{how does per-flow QoS
degrade as more TSN endpoints share a single gNB, and what determines
the practical capacity limit?}

This question is strongly affected by implementation choices that the
3GPP specifications leave open, especially the gNB MAC scheduler, which
allocates shared radio resources among competing traffic flows. Existing
work has compared schedulers for generic URLLC traffic~\cite{5gtq} and
studied time synchronisation over 5G in isolation~\cite{dasilva}, but no
prior study jointly characterises endpoint scaling, scheduler impact,
bandwidth sensitivity, fading behaviour, and gPTP stability under a
heterogeneous industrial TSN workload.

We address this gap using the nascTime simulation framework~\cite{nasctime-etfa}
on OMNeT++~6.3, INET~4.6, and Simu5G. The model includes closed-loop
control, machine vision, bulk telemetry, and IEEE~802.1AS/gPTP traffic
mapped through a four-bearer SDAP architecture. We evaluate 1 to~40
endpoints, four MAC schedulers---Maximum~C/I, Proportional Fair,
Deficit Round Robin, and QoS-aware Proportional Fair---two bandwidths
(\SI{10}{\mega\hertz} and \SI{20}{\mega\hertz}), and both ideal and
Jakes fading channels.

The results show three operating regimes. Below saturation, non-DRR
schedulers provide similar performance and scheduler choice has limited
impact. Near saturation, QoS-aware scheduling reduces critical-flow P99
latency by one to two orders of magnitude relative to channel-aware or
fair scheduling. Under overload, only QoS-aware scheduling maintains
near-complete delivery for the highest-priority traffic class, while
other schedulers exhibit spatial unfairness or class starvation. Across
the two evaluated bandwidths, the saturation threshold approximately
doubles when bandwidth increases from \SI{10}{\mega\hertz} to
\SI{20}{\mega\hertz}.

The study also exposes two architectural risks. First, the evaluated
per-UE DRR implementation is unsuitable for heterogeneous TSN bridging:
although it preserves closed-loop-control delivery, it starves
machine-vision and telemetry flows because its fixed per-user quantum is
not flow-class aware. Second, IEEE~802.1AS/gPTP traffic requires
explicit bearer treatment. A dedicated high-priority gPTP bearer avoids
the clock-servo divergence observed when synchronisation traffic
competes on the default bearer, but it does not fully bound residence
time under saturation because endpoints carrying lower-priority data
receive fewer MAC scheduling opportunities.

This paper makes three contributions:
\begin{enumerate}
    \item It characterises the scalability of 5G-TSN bridges under a
    heterogeneous industrial traffic mix, identifying three operating
    regimes and showing that, for the evaluated workload, the saturation
    threshold approximately doubles when bandwidth doubles.

    \item It evaluates a four-bearer SDAP architecture with a dedicated
    high-priority IEEE~802.1AS/gPTP bearer, showing that bearer
    isolation reduces clock-servo instability but does not by itself
    guarantee bounded gPTP residence time under saturation.

    \item It compares scheduler behaviour under ideal and Jakes fading
    channels, showing that the relative scheduler ranking is preserved
    in the tested configurations and motivating broader study of fading
    effects in 5G-TSN bridge deployments.
\end{enumerate}

The remainder of the paper is organised as follows.
Section~\ref{sec:background} reviews 5G-TSN bridge architecture and
related work. Section~\ref{sec:architecture} presents the four-bearer
SDAP design. Section~\ref{sec:methodology} describes the simulation
methodology. Section~\ref{sec:results} reports scaling, bandwidth,
fading, and synchronisation results. Section~\ref{sec:discussion}
discusses deployment implications and limitations, and
Section~\ref{sec:conclusion} concludes.

\section{Background and Related Work}\label{sec:background}

\subsection{5G-TSN Bridge Architecture}\label{sec:bg-arch}

3GPP Release~16 specifies a 5G system as a logical IEEE~802.1
TSN bridge between wired TSN segments~\cite{3gpp23501}. The bridge
is formed by a Network-side TSN Translator~(NW-TT), co-located with
the UPF, and a Device-side TSN Translator~(DS-TT), co-located with
the UE. The TSN Application Function exposes bridge capabilities such
as port delay, supported rates, and traffic-class support to the TSN
Centralised Network Controller.

QoS mapping follows a multi-stage pipeline. At the NW-TT, the
IEEE~802.1Q Priority Code Point~(PCP) is mapped to an IPv4 DSCP value;
the UPF Traffic Flow Filter maps DSCP to a 5G QoS Flow Identifier~(QFI);
and the SDAP layer maps each QFI to a Data Radio Bearer~(DRB). Each DRB
carries QoS attributes, including GBR flag, packet delay budget, packet
error rate, and priority level, which the gNB scheduler may use but is
not required to use. At the DS-TT, the reverse mapping reconstructs the
original TSN priority.

For time synchronisation, the 5G system implements IEEE~802.1AS
transparent-clock behaviour~\cite{ieee8021as}. The NW-TT timestamps
each gPTP Sync message at ingress, while the DS-TT computes the 5G
residence time and adds it to the gPTP correction field. This mechanism
is effective only if residence time and its variance remain bounded.
Thus, although the 3GPP bridge specifies the QoS and synchronisation
interfaces, the MAC scheduler ultimately determines how shared radio
resources are allocated among DRBs and endpoints. This implementation
choice is the source of the performance variability studied here.

\subsection{MAC Scheduling Disciplines}\label{sec:bg-sched}

We evaluate four Simu5G scheduling disciplines~\cite{simu5g}, chosen
to represent distinct radio-resource allocation strategies.

\paragraph{Maximum C/I (MaxCI)}
MaxCI assigns resource blocks to the UE with the highest instantaneous
CQI. It maximises instantaneous throughput but can create spatial
unfairness because UEs with persistently weaker channels receive fewer
resources. MaxCI is unaware of DRB QoS attributes and treats all traffic
within a UE identically.

\paragraph{Proportional Fair (PF)}
PF schedules the UE that maximises the ratio of instantaneous achievable
rate to exponentially averaged past rate~\cite{pf-scheduler}. This
balances throughput and fairness, but like MaxCI, it does not use
DRB-level QoS attributes and therefore cannot distinguish traffic
classes within a UE.

\paragraph{Deficit Round Robin (DRR)}
DRR allocates a fixed quantum of resource blocks to each UE in cyclic
order, carrying unused deficit to later rounds~\cite{drr-shreedhar}.
It provides strict per-UE fairness and is channel-unaware. In the
evaluated implementation, DRR does not distinguish among DRBs within a
UE, so control, vision, telemetry, and gPTP traffic compete within the
same per-user allocation.

\paragraph{QoS-aware Proportional Fair (QoS-PF)}
QoS-PF extends PF by weighting each connection according to DRB
attributes. In the Simu5G implementation, the weight for DRB~$d$ is
\begin{equation}\label{eq:qos-weight}
w_d =
(\text{GBR}_d\;?\;2:1)
\times
\frac{10}{p_d + 1}
\times
b(D_d),
\end{equation}
where $p_d$ is the priority level, $D_d$ is the delay budget, and
$b(D_d) \in \{5.0, 3.0, 1.5\}$ for delay budgets
$\leq 10$, $\leq 50$, and $\leq 100$\,ms, respectively. The scheduling
score is $s = w_d r_{\text{inst}}/r_{\text{avg}}$, using the same
instantaneous and averaged rates as PF. QoS-PF is therefore the only
evaluated scheduler that can explicitly prioritise high-priority DRBs
with tight delay budgets. Further details are provided in~\cite{qos-pf}.

\subsection{Related Work}\label{sec:bg-related}

Several simulation frameworks study 5G-TSN integration, but they differ
substantially in radio fidelity, QoS mapping, scaling scope, and
synchronisation support. 5GTQ~\cite{5gtq} implements NW-TT and DS-TT
modules in OMNeT++/Simu5G with QoS-aware scheduling, but maps traffic
priority directly at the MAC using 5QI values rather than SDAP-based
per-flow DRB selection; it also does not model gPTP residence time or
scaling beyond a small number of traffic classes. Da~Silva
et al.~\cite{dasilva} extend INET's IEEE~802.1AS model to transport
gPTP over 5G and demonstrate sub-microsecond synchronisation accuracy,
but do not include data-plane forwarding and therefore cannot capture
contention between TSN data traffic and synchronisation traffic.
6GDetCom~\cite{6gdetcom} models deterministic communication across
heterogeneous domains, but represents the wireless segment as a
statistical delay element rather than a 5G RAN with MAC scheduling,
HARQ, channel models, and resource-block allocation. P5G-TSN~\cite{p5gtsn}
adds TDD-pattern and resource-allocation analysis for private 5G, but
does not provide SDAP-based per-flow DRB mapping.

\begin{table}[t]
\centering
\caption{Comparison with existing 5G-TSN simulation studies.
\checkmark\,=\,supported, \texttimes\,=\,not supported,
$\circ$\,=\,partial.}
\label{tab:related-comparison}
\footnotesize
\resizebox{\linewidth}{!}{
\begin{tabular}{@{}lcccccc@{}}
\toprule
                & 5G$^a$  & SDAP/$^b$ & Sched.$^c$  & Scale$^d$  & gPTP$^e$  & Fading$^f$ \\
Study           & radio & DRB   & compare & sweep  & sync  &        \\
\midrule
5GTQ~\cite{5gtq}
    & \checkmark & \texttimes & $\circ$ & \texttimes & \texttimes & \texttimes \\
Da~Silva~\cite{dasilva}
    & \checkmark & \texttimes & \texttimes & \texttimes & \checkmark & \texttimes \\
6GDetCom~\cite{6gdetcom}
    & \texttimes & \texttimes & \texttimes & \texttimes & \texttimes & \texttimes \\
P5G-TSN~\cite{p5gtsn}
    & \checkmark & \texttimes & $\circ$ & \texttimes & \texttimes & $\circ$ \\
Zanbouri~\cite{zanbouri-scalability}
    & \checkmark & \texttimes & \texttimes & \checkmark & \texttimes & \checkmark \\
CQDRR~\cite{cqdrr}
    & $\circ$    & \texttimes & \checkmark & \texttimes & \texttimes & \checkmark \\
nascTime~\cite{nasctime-etfa}
    & \checkmark & \checkmark & \texttimes & \texttimes & \checkmark & \checkmark \\
\midrule
\textbf{This work}
    & \checkmark & \checkmark & \checkmark & \checkmark & \checkmark & \checkmark \\
\bottomrule
\multicolumn{7}{@{}p{0.95\columnwidth}@{}}{\scriptsize
$^a$Includes MAC scheduler, HARQ, channel model, RB allocation.
$^b$Per-flow QFI-to-DRB mapping via SDAP.
$^c$Systematic comparison of $\geq$3 scheduling disciplines.
$^d$Endpoint count swept from single-UE to cell-capacity limit.
$^e$gPTP transport through simulated radio with residence-time measurement.
$^f$Small-scale fading model evaluated.}
\end{tabular}
}
\end{table}

The nascTime framework~\cite{nasctime-etfa}, used in this work,
implements the full PCP$\to$DSCP$\to$QFI$\to$DRB pipeline and transports
gPTP through the simulated 5G radio path using L2-in-GTP-U encapsulation
with per-message residence-time correction. Its initial evaluation,
however, considers only a three-endpoint topology and does not study
endpoint scaling or scheduler choice.

Prior scalability and scheduling studies are complementary but narrower
than the present work. Zanbouri et al.~\cite{zanbouri-scalability}
evaluate bounded delay as the number of factory devices increases, but
do not compare MAC schedulers or study synchronisation under data-plane
contention. Ginth\"or et al.~\cite{ginthor} analyse multi-user scheduling
for industrial 5G, but consider a single traffic class and do not use
the SDAP/DRB QoS mapping pipeline. Zhang et al.~\cite{cqdrr} propose a
channel- and queue-aware DRR variant for hybrid TSN-5G traffic; their
focus is scheduler design, whereas we characterise existing schedulers
under heterogeneous scaling. Larrañaga et al.~\cite{larranaga} coordinate
configured grants with TSN scheduling, addressing a different resource
allocation mechanism from the dynamic schedulers evaluated here.

Table~\ref{tab:related-comparison} summarises the gap. To our knowledge,
no prior work jointly provides actual 5G radio simulation with MAC
scheduling and channel models, SDAP-based per-flow DRB mapping,
comparison of multiple schedulers, endpoint scaling to cell-capacity
limits, gPTP analysis under data-plane contention, and both bandwidth
and fading evaluation.

\section{System Architecture}\label{sec:architecture}

\begin{figure}[!t]
\centering
\includegraphics[width=0.9\columnwidth]{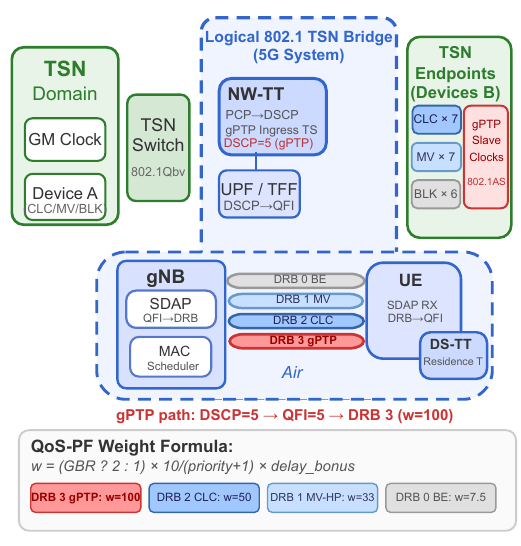}
\caption{5G-TSN bridge architecture with four-bearer SDAP configuration.
DRB~3 carries gPTP traffic at priority~0. QoS-PF weights are shown for
each DRB.}
\label{fig:arch}
\end{figure}

The 3GPP bridge specification defines the QoS mapping pipeline and
transparent-clock mechanism, but not how many Data Radio Bearers~(DRBs)
to provision, how traffic classes should be assigned to them, or whether
the synchronisation plane should receive dedicated radio resources.
These choices determine how the MAC scheduler observes and prioritises
competing TSN flows. This section describes the bearer architecture and
gPTP transport, shown in Fig. \ref{fig:arch}, which are used in our evaluation.

\subsection{Four-Bearer SDAP Configuration}\label{sec:four-drb}

We configure four DRBs per UE, each carrying a distinct traffic class
with QoS attributes matched to its TSN requirements
(Table~\ref{tab:drb-mapping}).

Closed-loop control~(CLC) has the tightest deadline
(\SI{2}{\milli\second}) and is assigned to DRB~2 with GBR treatment and
priority~1. Machine-vision high-priority traffic~(MV-HP) uses DRB~1
with a \SI{10}{\milli\second} delay budget. Machine-vision best-effort
traffic~(MV-BE) and bulk telemetry~(BLK) share DRB~0 because their
deadlines are looser and they do not require GBR service. gPTP traffic
uses the dedicated DRB~3.

The TSN-to-5G QoS mapping is a pass-through pipeline:
PCP~7/6/0 are mapped at the NW-TT to DSCP~7/6/0 for CLC, MV-HP, and
best-effort traffic, respectively; the UPF Traffic Flow Filter maps
DSCP~$n$ to QFI~$n$; and SDAP routes each QFI to its configured DRB.
This preserves TSN priority semantics without complex UPF policy rules.

Under QoS-PF, Eq.~\eqref{eq:qos-weight} gives a priority hierarchy:
DRB~3~(gPTP) receives weight~100, followed by DRB~2~(CLC) at~50,
DRB~1~(MV-HP) at~33.3, and DRB~0~(best effort) at~7.5. Under MaxCI, PF,
and DRR, these weights are ignored: all DRBs within a UE are treated
identically, with allocation driven by channel quality or cyclic order.
Thus, the bearer configuration provides the mechanism for QoS
differentiation, but only QoS-PF uses it. The experiments quantify the
consequence of this scheduler--bearer asymmetry as endpoint count
increases.

\begin{table}[!t]
\centering
\caption{SDAP bearer configuration.}
\label{tab:drb-mapping}
\begin{tabular}{@{}clccccr@{}}
\toprule
DRB & Traffic class & QFI & GBR & Delay & Priority & $w_d$ \\
    &               &     &     & (ms)  & (0\,=\,high) & \\
\midrule
0 & Best effort       & 0 & No  & 50  & 3 &   7.5 \\
  & (MV-BE, BLK)      &   &     &     &   &       \\
1 & MV high-priority  & 6 & Yes & 10  & 2 &  33.3 \\
2 & Closed-loop ctrl  & 7 & Yes &  2  & 1 &  50.0 \\
3 & gPTP (dedicated)  & 5 & Yes &  1  & 0 & 100.0 \\
\bottomrule
\end{tabular}
\end{table}

\subsection{Dedicated gPTP Bearer}\label{sec:gptp-bearer}

In the baseline 3GPP bridge path, gPTP frames share the user plane with
data traffic. When encapsulated in GTP-U for Ethernet-type PDU sessions,
the resulting IP packets carry no QoS marking by default and are routed
to the best-effort bearer~(DRB~0). In initial scaling experiments
without a dedicated bearer, gPTP frames at $N=20$ and
\SI{10}{\mega\hertz} were delayed by data-plane contention; the
IEEE~802.1AS slave clock servo then overcorrected, and in several runs
the INET SettableClock oscillator compensation exceeded its valid range
($[-500{,}000,\;+1{,}000{,}000]$~ppm), terminating the simulation. This
occurred under both PF and QoS-PF, indicating an architectural rather
than scheduler-specific vulnerability.

We therefore assign gPTP to DRB~3. The NW-TT marks GTP-U-encapsulated
gPTP packets with DSCP~5, the UPF maps DSCP~5 to QFI~5, and SDAP routes
QFI~5 to DRB~3, configured with GBR service, a
\SI{1}{\milli\second} delay budget, packet error rate $10^{-5}$, and
priority~0. Under QoS-PF this yields weight~100, so gPTP is scheduled
ahead of data traffic whenever its UE receives a scheduling opportunity.

To retain failed cases for analysis, we modified INET SettableClock to
clamp out-of-range oscillator-compensation values rather than terminate
the simulation. This instrumentation does not affect normal operation;
it records clock-servo stress in cases that would otherwise diverge. In
the primary sweep with the dedicated bearer, no clamping events occur at
$N\leq10$; events appear only under deep saturation
($N=20$, \SI{10}{\mega\hertz}) for endpoints with reduced scheduling
frequency, as analysed in Section~\ref{sec:res-gptp}.

The dedicated bearer provides \emph{intra-UE} priority but not
\emph{inter-UE} scheduling guarantees. gPTP is highest priority within a
scheduled UE, but a UE carrying only low-priority data may receive fewer
MAC scheduling turns, delaying its gPTP frames despite DRB~3 priority.
This coupling is evaluated in Section~\ref{sec:results}. The capacity
cost is small: gPTP generates one Sync/FollowUp pair every
\SI{125}{\milli\second}, each approximately 100~bytes, and the extra DRB
uses one of the 29 DRBs permitted per PDU session~\cite{3gpp38300}.

\section{Experimental Methodology}\label{sec:methodology}

\subsection{Simulation Platform}\label{sec:platform}

Experiments use the nascTime framework~\cite{nasctime-etfa}, built on
OMNeT++~6.3, INET~4.6.0, and Simu5G~v1.4.1~\cite{sdap}. nascTime
implements NW-TT and DS-TT modules integrated with INET's
\texttt{LayeredEthernetInterface} and transports gPTP over the simulated
5G path using L2-in-GTP-U encapsulation with per-message residence-time
measurement, as described in Section~\ref{sec:gptp-bearer}.

The topology contains a TSN grandmaster, TSN switch, NW-TT connected to
the 5G UPF, one gNB, $N$ UEs each co-located with a DS-TT, and $N$
downstream TSN endpoints. An upstream TSN device~(Device~A) hosts all
traffic sources; each UE maps one-to-one to a downstream endpoint
(Device~B).

\subsection{Radio Configuration}\label{sec:radio-config}

Table~\ref{tab:ran-config} lists the 5G~NR parameters. We evaluate
\SI{10}{\mega\hertz} and \SI{20}{\mega\hertz} bandwidths, corresponding
to 25 and 51 RBs, respectively, in band~n78 at \SI{3.5}{\giga\hertz}.
Both use numerology $\mu=1$ (\SI{30}{\kilo\hertz} SCS), giving a
\SI{0.5}{\milli\second} slot duration. The primary experiments disable
fading and shadowing to isolate scheduling and contention effects; the
fading sweep enables six-path Jakes fading.

\begin{table}[t]
\centering
\caption{5G NR radio configuration.}
\label{tab:ran-config}
\begin{tabular}{@{}ll@{}}
\toprule
Parameter & Value \\
\midrule
Carrier frequency      & \SI{3.5}{\giga\hertz} (band n78) \\
Numerology ($\mu$)     & 1 (\SI{30}{\kilo\hertz} SCS) \\
Slot duration          & \SI{0.5}{\milli\second} \\
Bandwidth A            & 25 RBs ($\approx$\SI{10}{\mega\hertz}) \\
Bandwidth B            & 51 RBs ($\approx$\SI{20}{\mega\hertz}) \\
Channel model          & NrChannelModel (Indoor Hotspot) \\
Fading                 & Disabled; Jakes, 6 paths in fading sweep \\
Shadowing              & Disabled \\
HARQ                   & Enabled (default Simu5G config.) \\
\bottomrule
\end{tabular}
\end{table}

\subsection{Traffic Model and Endpoint Scaling}\label{sec:traffic-model}

The workload comprises closed-loop control~(CLC), machine vision~(MV),
and bulk telemetry~(BLK), derived from 3GPP TR~22.804 service
requirements~\cite{3gpp22804}. Table~\ref{tab:traffic} gives packet
sizes, periods, deadlines, and DRB mappings.

\begin{table}[t]
\centering
\caption{Traffic profiles and TSN flow parameters.}
\label{tab:traffic}
\begin{tabular}{@{}llcccc@{}}
\toprule
Profile & Flow & Size & Period & Deadline & DRB \\
        &      & (B)  &        & (ms)     &     \\
\midrule
CLC & clc\_hp & 100 & \SI{1}{\milli\second} CBR & 2 & 2 \\
\midrule
\multirow{2}{*}{MV}
    & mv\_hp  & 1500 & \SI{5}{\milli\second} CBR & 10 & 1 \\
    & mv\_be  & 500  & exp(\SI{2}{\milli\second}) & 50 & 0 \\
\midrule
BLK & blk\_be & 1200 & exp(\SI{10}{\milli\second}) & 100 & 0 \\
\bottomrule
\end{tabular}
\end{table}

CLC models a \SI{1}{\milli\second} periodic control loop and is mapped
to DRB~2. MV contains a high-priority periodic stream on DRB~1 and a
best-effort background stream on DRB~0. BLK models lower-priority
telemetry on DRB~0. Each endpoint also generates a reverse-direction CBR
flow of 100~B every \SI{10}{\milli\second}
(\SI{0.08}{\mega\bit\per\second}), excluded from the downlink load
totals.

Endpoint counts range from $N=1$ to~40 using the profile mix in
Table~\ref{tab:endpoint-mix}. Offered downlink load is computed as
$L=8S/T$, where $S$ is packet size and $T$ is the fixed period or mean
inter-arrival time. Per-endpoint downlink loads are
\SI{0.8}{\mega\bit\per\second} for CLC,
\SI{4.4}{\mega\bit\per\second} for MV
(MV-HP plus MV-BE), and \SI{0.96}{\mega\bit\per\second} for BLK.

\begin{table}[t]
\centering
\caption{Endpoint profile mix and aggregate downlink offered load
per $N$. Each MV endpoint generates both MV-HP and MV-BE flows.}
\label{tab:endpoint-mix}
\begin{tabular}{@{}rcccc@{}}
\toprule
$N$ & CLC & MV & BLK & Offered DL load \\
    &     &    &     & (Mbps) \\
\midrule
 1 &  1 &  0 &  0 &   0.8 \\
 5 &  2 &  2 &  1 &  11.4 \\
10 &  4 &  4 &  2 &  22.7 \\
15 &  5 &  5 &  5 &  30.8 \\
20 &  7 &  7 &  6 &  42.2 \\
30 & 10 & 10 & 10 &  61.6 \\
40 & 14 & 14 & 12 &  84.3 \\
\bottomrule
\end{tabular}
\end{table}

At \SI{10}{\mega\hertz}, effective downlink capacity is approximately
\SI{25}{\mega\bit\per\second}; therefore $N=10$ approaches capacity,
while $N=15$ and $N=20$ exceed it by approximately 23\% and 69\%,
respectively. At \SI{20}{\mega\hertz}, effective capacity is
approximately \SI{50}{\mega\bit\per\second}; $N=20$ remains below
capacity, while $N=30$ exceeds it by approximately 23\%.

The protocol stack uses Simu5G~v1.4.1 defaults unless stated otherwise:
RLC UM, HARQ with up to three retransmissions, FDD duplexing, and
\SI{2}{\mega\byte} FIFO MAC buffers per logical channel
(per DRB per UE). No application-layer deadline dropping is configured:
packets remain queued until delivered, lost through RLC buffer overflow,
or the simulation ends. Configuration files and analysis scripts are
available in the nascTime repository~\cite{scalability-nasc}.

\subsection{Experiment Matrix}\label{sec:experiment-matrix}

Each simulation runs for \SI{30}{\second} with a
\SI{2}{\second} warmup; metrics are collected over
$t=\SI{2}{\second}$ to $t=\SI{30}{\second}$. Each configuration uses
three random seeds. Table~\ref{tab:sweeps} summarises the four sweeps.

The primary sweeps compare MaxCI, PF, DRR, and QoS-PF across all
endpoint counts and both bandwidths. The fading sweeps compare MaxCI
and QoS-PF under ideal and Jakes channels to test whether fading alters
the scheduler ranking. gPTP residence time and clock-servo data are
collected in both primary and fading sweeps. A fifth Simu5G scheduler,
MAXCI\_COMP, was tested initially but produced results identical to
MaxCI and is excluded.

\subsection{Metrics and Packet Accounting}\label{sec:metrics}

We report five metrics. Packet Delivery Ratio~(PDR) is the number of
unique application-layer packets received at Device~B divided by the
number generated at Device~A, computed per flow class. End-to-end delay
is measured from packet creation at Device~A to reception at Device~B
using the receiver-side \texttt{endToEndDelay} vector; we report mean,
P99, P99.9, and maximum. Deadline compliance is the fraction of
(endpoint, seed) cells whose P99 delay does not exceed the flow's TSN
deadline. Spatial fairness is the standard deviation of PDR across
endpoints for a given $(N,\text{scheduler},\text{flow class})$,
averaged over seeds. gPTP residence time is extracted from the DS-TT
\texttt{residenceTime} vector, and clock-servo stability from each
Device~B clock's \texttt{oscillatorCompensationChanged} vector.

All metrics use the measurement window from $t=\SI{2}{\second}$ to
$t=\SI{30}{\second}$. The PDR denominator is the number of unique
application-layer packets generated in this window; the numerator is the
number of matching unique packets received at the destination, whether
or not they meet their TSN deadline. Delay statistics include all
delivered packets, including late packets. Deadline compliance is
therefore reported separately to distinguish ``delivered eventually''
from ``delivered within deadline.'' Packets generated before the
measurement window are excluded from both numerator and denominator.

\begin{table}[t]
\centering
\caption{Experiment sweeps.}
\label{tab:sweeps}
\begin{tabular}{@{}lllcr@{}}
\toprule
Sweep & BW & $N$ values & Schedulers & Cells \\
\midrule
Primary & \SI{10}{\mega\hertz} & 1, 5, 10, 15, 20 & 4 & 60 \\
Primary & \SI{20}{\mega\hertz} & 1, 5, 10, 15, 20, 30, 40 & 4 & 84 \\
Fading  & \SI{10}{\mega\hertz} & 1, 10, 20 & 2$^{\dagger}$ & 36 \\
Fading  & \SI{20}{\mega\hertz} & 1, 10, 20 & 2$^{\dagger}$ & 36 \\
\midrule
\multicolumn{4}{@{}l}{Total simulation cells} & 216 \\
\bottomrule
\multicolumn{5}{@{}l}{\footnotesize $^{\dagger}$MaxCI and QoS-PF only,
each with fading$\,\in\,$\{off, on\}.}
\end{tabular}
\end{table}

\section{Results}\label{sec:results}

We present results as a scaling narrative: no-contention baseline,
saturation onset, overload, bandwidth sensitivity, fading resilience,
and time synchronisation. The analysis focuses on MaxCI, PF, DRR, and
QoS-PF; MAXCI\_COMP is excluded as noted in
Section~\ref{sec:experiment-matrix}. The comparison should be read as
evaluating the benefit of exposing TSN class information to the MAC
scheduler, not as a universal ranking of scheduling algorithms.

%%--------------------------------------------------------------------
\subsection{Baseline: The Latency Floor}\label{sec:res-floor}

At $N=1$, a single CLC endpoint operates without contention. All four
schedulers deliver 28{,}000 packets, i.e., 100\% of the offered load,
with mean and P99 delay of \SI{2.25}{\milli\second}. This effective
latency floor arises from the \SI{0.5}{\milli\second} slot duration at
$\mu=1$, scheduling-request/grant processing, and PDCP/RLC handling.

Because the CLC deadline is \SI{2}{\milli\second}, the evaluated
sub-6\,GHz, \SI{30}{\kilo\hertz}-SCS configuration does not satisfy
strict CLC deadline compliance even under zero contention. Subsequent
scheduler comparisons therefore measure \emph{relative robustness under
load}, not strict TSN deadline satisfaction. Meeting deadlines below
approximately \SI{3}{\milli\second} would require radio-configuration
changes such as configured grants, pre-emptive scheduling, or higher
numerology.

\begin{figure}[t]
\centering
\includegraphics[width=0.95\columnwidth]{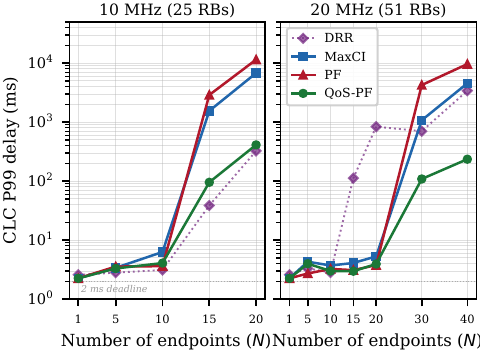}
\caption{CLC P99 end-to-end delay vs.\ endpoint count $N$ at
\SI{10}{\mega\hertz} and \SI{20}{\mega\hertz}. Logarithmic $y$-axis.
The \SI{2}{\milli\second} deadline is shown as a dotted reference.
DRR is omitted from the \SI{10}{\mega\hertz} panel at $N \geq 15$ to
preserve readability.}
\label{fig:clc-p99}
\end{figure}

%%--------------------------------------------------------------------
\subsection{Scaling at \SI{10}{\mega\hertz}}\label{sec:res-10mhz}

Figure~\ref{fig:clc-p99} shows CLC P99 delay versus endpoint count.
At \SI{10}{\mega\hertz}, three regimes emerge.

\subsubsection{Below Saturation ($N \leq 10$)}

At $N=5$, all non-DRR schedulers deliver 100\% of all traffic classes.
CLC P99 ranges from \SI{3.31}{\milli\second} (QoS-PF) to
\SI{3.56}{\milli\second} (PF), and MV-HP delivery is 100\%. BLK delivery
ranges from 2{,}818 (PF) to 2{,}895 (MaxCI) out of approximately
2{,}800 offered, reflecting the variance of the exponential arrival
process. Scheduler choice has limited impact in this regime.

At $N=10$, CLC delivery remains 100\% for all non-DRR schedulers, with
P99 ranging from \SI{3.62}{\milli\second} (PF) to
\SI{6.29}{\milli\second} (MaxCI). PF achieves the lowest CLC latency at
this operating point, while MV-HP delivery remains 100\%.

DRR already shows its characteristic pathology. At $N=5$, it delivers
only 121 of 5{,}600 MV-HP packets (2.2\%) and 215 of approximately
2{,}800 BLK packets (7.7\%), despite the cell operating below capacity.
MV-HP and BLK mean delays reach \SI{4.4}{\second} and
\SI{9.4}{\second}, respectively, while CLC remains at 100\% delivery
with \SI{2.33}{\milli\second} mean delay. This starvation is structural:
the fixed per-UE quantum accommodates CLC's
\SI{800}{\kilo\bit\per\second} CBR flow but cannot serve burstier MV and
BLK traffic within a single round-robin turn.

\subsubsection{Saturation Onset ($N = 15$)}

Between $N=10$ and $N=15$, offered load increases from
\SI{22.7}{\mega\bit\per\second} to
\SI{30.8}{\mega\bit\per\second}, exceeding the effective
\SI{25}{\mega\bit\per\second} cell capacity by approximately 23\%.
Scheduler rankings then diverge sharply.

QoS-PF maintains CLC delivery at 99.9\% (27{,}984 of 28{,}000), with
mean delay of \SI{16}{\milli\second} and P99 of
\SI{96}{\milli\second}. MaxCI delivers 95.2\% of CLC (26{,}658) with
P99 of \SI{1.55}{\second}; PF delivers 89.3\% (25{,}010) with P99 of
\SI{2.89}{\second}. Thus, at saturation onset, QoS-PF reduces CLC P99
delay by one to two orders of magnitude relative to MaxCI and PF.

Spatial fairness explains this divergence. The standard deviation of
CLC delivery across endpoints is 31 packets for QoS-PF, compared with
1{,}873 for MaxCI and 2{,}233 for PF. MaxCI favours endpoints with
better channel quality, while PF distributes resources without
flow-class awareness and therefore does not prioritise CLC. MV-HP
delivery also begins to separate: QoS-PF delivers 98.0\% (5{,}489 of
5{,}600), MaxCI 86.6\% (4{,}851), and PF 87.4\% (4{,}894).

\begin{figure}[t]
\centering
\includegraphics[width=0.9\columnwidth]{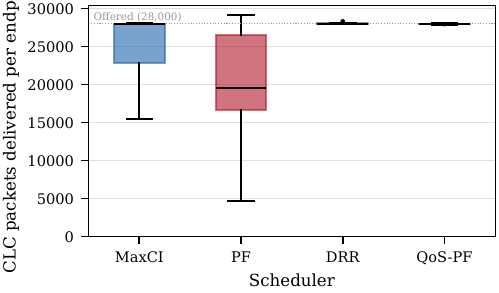}
\caption{Distribution of per-endpoint CLC delivery at $N=20$,
\SI{10}{\mega\hertz}. Each box spans the interquartile range across
endpoints and repetitions. MaxCI and PF exhibit high variance; QoS-PF
and DRR maintain tight distributions.}
\label{fig:fairness-boxplot}
\end{figure}

\begin{table}[t]
\centering
\caption{Performance at $N = 20$, \SI{10}{\mega\hertz}. All delivery
counts are \emph{mean per-endpoint} values averaged across repetitions. PDR is computed per endpoint and averaged. Fairness is the standard deviation of
per-endpoint delivery across endpoints.}
\label{tab:n20-summary}
\begin{tabular}{@{}lrrrr@{}}
\toprule
Metric & MaxCI & PF & DRR$^\dagger$ & QoS-PF \\
\midrule
CLC delivery        & 21{,}698 & 16{,}538 & 28{,}084 & 27{,}955 \\
CLC PDR (\%)        & 77.5     & 59.1     & 100.0    & 99.8     \\
CLC mean (ms)       & 3{,}634  & 4{,}923  & 78       & 54       \\
CLC P99 (ms)        & 6{,}790  & 11{,}472 & 324      & 412      \\
CLC fairness (std)  & 5{,}756  & 6{,}961  & 54       & 54       \\
\midrule
MV-HP delivery      & 3{,}142  & 3{,}549  & 120      & 3{,}074  \\
MV-HP PDR (\%)      & 56.1     & 63.4     & 2.1      & 54.9     \\
BLK delivery        & 2{,}797  & 1{,}929  & 214      & 2{,}382  \\
BLK PDR (\%)        & 99.9     & 68.9     & 7.6      & 85.1     \\
\bottomrule
\multicolumn{5}{@{}p{0.95\columnwidth}@{}}{\scriptsize
$^\dagger$DRR CLC delivery count exceeds the nominal offered count due
to measurement-window boundary effects; PDR is capped at 100.0\%.}
\end{tabular}
\end{table}

\subsubsection{Overload ($N = 20$)}\label{sec:res-overload}

At $N=20$, offered load substantially exceeds capacity and scheduler
differences become stark. Table~\ref{tab:n20-summary} summarises the
main metrics.

QoS-PF delivers 99.8\% of CLC traffic with P99 delay of
\SI{412}{\milli\second} and delivery standard deviation of 54 packets.
PF delivers only 59.1\% of CLC with P99 of \SI{11.5}{\second} and
fairness standard deviation of 6{,}961, indicating severe endpoint-level
imbalance. Figure~\ref{fig:fairness-boxplot} shows the same pattern:
MaxCI and PF have wide delivery distributions, while QoS-PF and DRR
remain tight.

\begin{figure}[t]
\centering
\includegraphics[width=0.9\columnwidth]{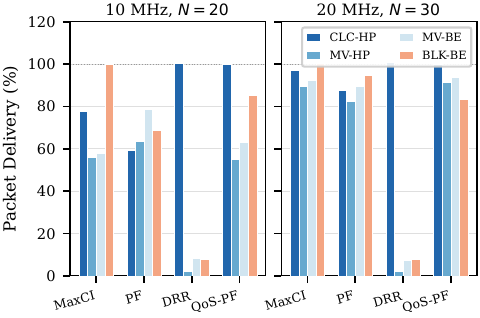}
\caption{Packet delivery ratio per flow class under saturation.
Left: \SI{10}{\mega\hertz}, $N=20$. Right: \SI{20}{\mega\hertz},
$N=30$. DRR delivers CLC at 100\% but starves MV-HP and BLK.}
\label{fig:delivery-saturation}
\end{figure}

DRR preserves the CLC stream almost completely in the measured window,
but only by starving other classes: MV-HP delivery is 2.1\% and BLK
delivery is 7.6\%. Figure~\ref{fig:delivery-saturation} visualises this
per-class delivery trade-off at the saturation points for both
bandwidths. Thus, the evaluated per-UE DRR implementation is unsuitable
for heterogeneous TSN bridging.

QoS-PF protects the highest-priority class through overload at the cost
of lower service to less critical flows. Its MV-HP delivery (54.9\%) is
below PF's (63.4\%), reflecting deliberate resource reallocation from MV
to CLC. This trade-off is appropriate only when the deployment priority
structure favours CLC preservation.

\begin{table}[t]
\centering
\caption{Cross-bandwidth comparison for CLC flow. Corresponding
operating points are paired by approximate load-to-capacity ratio.}
\label{tab:cross-bandwidth}
\begin{tabular}{@{}llrrrr@{}}
\toprule
              &           & \multicolumn{2}{c}{Saturation onset}
                          & \multicolumn{2}{c}{Overload} \\
\cmidrule(lr){3-4} \cmidrule(lr){5-6}
              &           & 10\,MHz    & 20\,MHz
                          & 10\,MHz    & 20\,MHz \\
              &           & $N{=}15$   & $N{=}30$
                          & $N{=}20$   & $N{=}40$ \\
\midrule
\multirow{3}{*}{PDR (\%)}
 & MaxCI  &  95.2 &  96.9 &  77.5 &  85.5 \\
 & PF     &  89.3 &  87.6 &  59.1 &  76.8 \\
 & QoS-PF &  99.9 &  99.9 &  99.8 &  100.0 \\
\midrule
\multirow{3}{*}{\shortstack[l]{P99\\(ms)}}
 & MaxCI  & 1{,}552 & 1{,}078 & 6{,}790 & 4{,}553 \\
 & PF     & 2{,}889 & 4{,}256 & 11{,}472 & 9{,}646 \\
 & QoS-PF &      96 &     109 &      412 &     236 \\
\midrule
\multirow{3}{*}{\shortstack[l]{Fairness\\(std)}}
 & MaxCI  & 1{,}873 & 1{,}965 & 5{,}756 & 4{,}942 \\
 & PF     & 2{,}233 & 3{,}282 & 6{,}961 & 6{,}412 \\
 & QoS-PF &      31 &      45 &      54 &      75 \\
\bottomrule
\end{tabular}
\end{table}

%%--------------------------------------------------------------------
\subsection{Bandwidth Sensitivity}\label{sec:res-bandwidth}

Figure~\ref{fig:clc-p99} (right panel) shows CLC P99 delay at
\SI{20}{\mega\hertz}. The same three-regime structure appears, but the
saturation threshold shifts to higher $N$.

At $N=20$, all non-DRR schedulers deliver 100\% of all traffic classes.
CLC P99 ranges from \SI{3.77}{\milli\second} (PF) to
\SI{5.30}{\milli\second} (MaxCI), comparable to the
\SI{10}{\mega\hertz}, $N=5$ case. PF again marginally outperforms
QoS-PF below saturation (\SI{3.77}{\milli\second} versus
\SI{3.94}{\milli\second}), suggesting that QoS weighting introduces a
small cost that is only justified under contention. This cost is modest
(typically $<$\SI{1.5}{\milli\second} in P99) and disappears near
saturation.

At $N=30$, saturation onset is visible. QoS-PF maintains 99.9\% CLC
delivery with P99 of \SI{109}{\milli\second}; MaxCI drops to 96.9\%
with P99 of \SI{1{,}078}{\milli\second}; and PF drops to 87.6\% with
P99 of \SI{4{,}256}{\milli\second}. CLC fairness standard deviation is
45 packets for QoS-PF, compared with 1{,}965 for MaxCI and 3{,}282 for
PF. At $N=40$, QoS-PF still delivers 99.96\% of CLC with P99 of
\SI{236}{\milli\second}, while MaxCI and PF fall to 85.5\% and 76.8\%
delivery with P99 delays of \SI{4{,}553}{\milli\second} and
\SI{9{,}646}{\milli\second}, respectively.

Across the two evaluated bandwidths, the saturation threshold
approximately doubles when bandwidth doubles. Table~\ref{tab:cross-bandwidth}
shows that $N=15$ at \SI{10}{\mega\hertz} and $N=30$ at
\SI{20}{\mega\hertz} produce comparable delivery, P99, and fairness
values; similarly, $N=20$ at \SI{10}{\mega\hertz} mirrors $N=40$ at
\SI{20}{\mega\hertz}. This proportional relationship can guide capacity
estimation for the evaluated workload, although more bandwidth points
would be needed to establish a general scaling law.
%%--------------------------------------------------------------------
\begin{figure}[t]
\centering
\includegraphics[width=0.9\columnwidth]{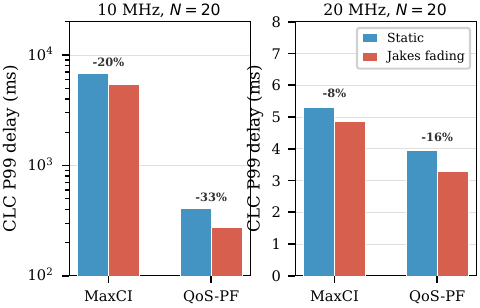}
\caption{Effect of Jakes fading on CLC P99 delay at $N=20$.
Left: \SI{10}{\mega\hertz} saturated case. Right:
\SI{20}{\mega\hertz} below-saturation case. Percentage labels show
change relative to the static channel.}
\label{fig:fading-impact}
\end{figure}

\subsection{Fading Resilience}\label{sec:res-fading}

Table~\ref{tab:fading} compares CLC P99 delay under static and Jakes
fading channels for MaxCI and QoS-PF.

\begin{table}[t]
\centering
\caption{CLC P99 delay (ms) under static and Jakes fading channels.}
\label{tab:fading}
\begin{tabular}{@{}rl rr rr@{}}
\toprule
  &  & \multicolumn{2}{c}{MaxCI}
     & \multicolumn{2}{c}{QoS-PF} \\
\cmidrule(lr){3-4} \cmidrule(lr){5-6}
BW & $N$ & Static & Fading & Static & Fading \\
\midrule
\multirow{3}{*}{\SI{10}{\mega\hertz}}
   &  1 &    2.25 &    2.25 &    2.25 &    2.25 \\
   & 10 &    6.29 &    5.29 &    4.08 &    3.58 \\
   & 20 & 6{,}790 & 5{,}431 &     412 &     275 \\
\midrule
\multirow{3}{*}{\SI{20}{\mega\hertz}}
   &  1 &    2.25 &    2.25 &    2.25 &    2.25 \\
   & 10 &    3.71 &    3.42 &    3.00 &    2.75 \\
   & 20 &    5.30 &    4.87 &    3.94 &    3.30 \\
\bottomrule
\end{tabular}
\end{table}

Two observations emerge. First, QoS-PF achieves lower CLC P99 than
MaxCI under every tested channel condition, so the scheduler ranking is
not contingent on ideal channels. Second, fading is associated with
lower latency in these experiments. At \SI{10}{\mega\hertz}, $N=20$,
QoS-PF P99 decreases from \SI{412}{\milli\second} to
\SI{275}{\milli\second} (33\%), and MaxCI decreases from
\SI{6{,}790}{\milli\second} to \SI{5{,}431}{\milli\second} (20\%).
At \SI{20}{\mega\hertz}, where the cell is below saturation, the effect
is smaller than 15\%.

This behaviour is consistent with multi-user diversity: under fading,
CQI varies across TTIs, allowing the scheduler to exploit favourable
channel instances. The effect is most visible under saturation, where
resource efficiency directly affects queueing.
Figure~\ref{fig:fading-impact} visualises this contrast at $N=20$:
the reduction is substantial in the saturated \SI{10}{\mega\hertz}
case and modest in the below-saturation \SI{20}{\mega\hertz} case.
However, this observation is based on one fading model (Jakes, six
paths), two bandwidths, and three endpoint counts. Broader channel,
mobility, and geometry studies are required before generalising the
result. CLC, MV-HP, and BLK delivery are unaffected by fading in the
tested configurations.

%%--------------------------------------------------------------------
\subsection{Time Synchronisation}\label{sec:res-gptp}

The dedicated gPTP bearer (DRB~3, Section~\ref{sec:gptp-bearer})
prevents the clock-servo divergence observed in initial experiments:
all simulation cells complete without oscillator-compensation errors.
However, gPTP residence time still degrades under saturation.

\subsubsection{Residence Time Under Scaling}

At $N=1$, 5G residence time is
$2{,}772 \pm 249\;\mu\text{s}$ (mean $\pm$ std), with maximum
\SI{3{,}000}{\micro\second}, and is stable across schedulers.

At $N=10$ without fading, residence time increases modestly. Under
MaxCI, mean per-DS-TT residence time ranges from
\SI{3{,}500}{\micro\second} to \SI{5{,}000}{\micro\second}, with
standard deviations of \SI{1{,}200}{\micro\second} to
\SI{2{,}500}{\micro\second}. Under QoS-PF, the range narrows to
\SI{3{,}000}{\micro\second}--\SI{3{,}500}{\micro\second}, with
standard deviations of \SI{230}{\micro\second}--\SI{1{,}080}{\micro\second}.
QoS-PF therefore produces 2--5$\times$ lower residence-time variance,
consistent with preferential scheduling of DRB~3.

At $N=20$ under saturation at \SI{10}{\mega\hertz}, residence times
diverge by endpoint type. CLC endpoints, which receive frequent
scheduling turns because of high-weight DRB~2 traffic, show residence
times of \SI{15}{\milli\second}--\SI{135}{\milli\second}. MV endpoints,
which carry lower-priority data and receive fewer scheduling turns,
show residence times of \SI{394}{\milli\second}--\SI{4{,}439}{\milli\second},
three orders of magnitude higher.

\subsubsection{The Inter-UE Scheduling Coupling}

This disparity shows that a dedicated bearer provides intra-UE priority
but not inter-UE scheduling guarantees. DRB~3 makes gPTP the
highest-priority traffic \emph{within} a scheduled UE. Under QoS-PF,
however, UEs carrying CLC on DRB~2 (weight~50) receive more scheduling
opportunities than UEs carrying MV or best-effort traffic on lower-weight
DRBs. As a result, gPTP residence time depends on the endpoint's data
traffic class, even though all endpoints use the same gPTP bearer.

Clock-servo data confirm this coupling. At $N=10$ under fading,
oscillator-compensation standard deviation is approximately
4{,}200~ppm, with corrections from $-$24{,}000 to $+$24{,}000~ppm.
At $N=20$ under saturation, MV-type endpoints exceed
190{,}000~ppm standard deviation, with peaks approaching the
$1{,}000{,}000$~ppm SettableClock clamping bound.

Thus, complete gPTP protection under saturation requires more than a
dedicated bearer: it also requires a minimum scheduling frequency per UE,
for example via semi-persistent scheduling or configured grants. We
return to this issue in Section~\ref{sec:discussion}.

\section{Discussion}\label{sec:discussion}

\subsection{Synthesis of Findings}

The results show that 5G-TSN bridge performance is governed by the
interaction between endpoint load, bearer architecture, and MAC
scheduling at a load-dependent operating point, rather than by endpoint
count or scheduler choice alone.

\paragraph{Scaling behaviour}
The transition from adequate to degraded performance is abrupt. At
\SI{10}{\mega\hertz}, CLC P99 latency increases by two orders of
magnitude between $N=10$ (\SI{4}{\milli\second}) and $N=15$
(\SI{96}{\milli\second} under QoS-PF; \SI{1.5}{\second} under MaxCI).
Thus, a deployment that performs acceptably below saturation may degrade
qualitatively after only a small increase in endpoints or offered load.
This sharp transition follows from fixed radio capacity: once offered
load approaches the scheduling limit, queues grow rapidly and
non-prioritised traffic is affected first.

\paragraph{Scheduler selection}
In the evaluated configuration, QoS-aware scheduling offers the best
risk trade-off. Below saturation, QoS-PF's cost relative to PF is small
($<$\SI{1.5}{\milli\second} in CLC P99); above saturation, it is the
only tested scheduler that maintains near-complete delivery for the
highest-priority traffic class. This supports using QoS-aware scheduling
for heterogeneous TSN traffic, while recognising that the result applies
to the tested Simu5G implementations.

The evaluated per-UE DRR implementation should not be used for
heterogeneous TSN bridging. Its per-UE fairness model cannot distinguish
a \SI{1}{\milli\second} control loop from bulk telemetry within the same
UE, causing starvation of bursty flows independently of utilisation.
Channel- or queue-aware DRR variants such as CQDRR~\cite{cqdrr} may
mitigate this limitation but are outside this evaluation.

\paragraph{Bandwidth planning}
Across the two evaluated bandwidths, the saturation threshold
approximately doubles when bandwidth doubles. For the workload studied
here, this corresponds to roughly $N=12$ per
\SI{10}{\mega\hertz}. This provides an initial planning heuristic, not a
general scaling law: additional bandwidths and traffic mixes would be
needed for broader validation.

\paragraph{Time synchronisation}
Isolating gPTP on a dedicated high-priority bearer prevented the
clock-servo divergence observed without it: at $N=20$, 4 of 15 cells
terminated without the bearer, while all cells completed with it.
However, bearer isolation alone does not bound gPTP residence time under
saturation. Endpoints carrying lower-priority data receive fewer MAC
scheduling turns, delaying their gPTP traffic despite its intra-UE
priority. This inter-UE scheduling coupling suggests that robust
synchronisation may require a minimum scheduling frequency per UE,
implemented through mechanisms such as semi-persistent scheduling,
configured grants, or priority handling before MAC scheduling. Evaluating
these mechanisms is left to future work.

\subsection{Deployment Guidelines}

The following guidelines are scoped to the evaluated simulation platform,
traffic mix, and scheduler implementations, and should be validated for
the target deployment:

\begin{enumerate}
    \item \textbf{Use or evaluate QoS-aware scheduling.} QoS-PF provides
    the strongest protection for critical traffic under saturation with
    negligible cost below saturation.

    \item \textbf{Provision a dedicated gPTP bearer.} A high-priority
    DRB for gPTP prevents the observed clock-servo divergence, although
    residence-time inflation can persist for scheduling-starved endpoints.

    \item \textbf{Plan around the saturation threshold.} The transition
    to overload is abrupt; deployments should characterise the threshold
    for their traffic mix and maintain margin below it.

    \item \textbf{Account for the radio latency floor.} In the evaluated
    grant-based \SI{30}{\kilo\hertz}-SCS Simu5G configuration, the
    effective floor is approximately \SI{2.25}{\milli\second}; tighter
    deadlines may require configured grants, pre-emptive scheduling, or
    higher numerology.

    \item \textbf{Monitor synchronisation per endpoint class.}
    Lower-priority endpoints can experience elevated gPTP residence time
    under saturation even with a dedicated bearer, so per-endpoint
    synchronisation monitoring is advisable.
\end{enumerate}

\subsection{Limitations}\label{sec:limitations}

Several limitations scope the applicability of these findings.

\paragraph{Single cell and no mobility}
All experiments use one gNB and stationary UEs. Multi-cell interference,
handover, coordinated scheduling, and mobility may change both the
saturation threshold and scheduler ranking.

\paragraph{Simulation fidelity}
Results reflect Simu5G's scheduler implementations, INET's protocol
stack, and the NrChannelModel channel model. Commercial gNBs may differ
in scheduling, buffering, HARQ, and queue management. The
\SI{2.25}{\milli\second} latency floor is therefore a property of this
configuration, not a universal 5G limit.

\paragraph{Traffic model}
The workload is a specific control/vision/telemetry mix derived from
3GPP TR~22.804. Correlated bursts, event-driven traffic, or different
uplink/downlink ratios may shift the operating regimes.

\paragraph{Fading scope}
The fading analysis uses one Jakes six-path model without mobility. The
observed latency reduction under fading may not generalise to channels
with deep fades, shadowing, mobility, or frequency selectivity.

\paragraph{No TSN-side scheduling}
The TSN domain uses FIFO forwarding without IEEE~802.1Qbv Time-Aware
Shaping or IEEE~802.1Qav Credit-Based Shaping. Coordinating TSN-side and
5G scheduling remains outside the scope of this study.

\paragraph{Statistical scope}
Each configuration uses three seeds. This supports the main trends but
limits strong claims about P99.9 and bursty-tail behaviour; additional
repetitions and confidence intervals would strengthen those claims.

\paragraph{Scaling confounds}
The scaling sweep varies endpoint count, offered load, and traffic-class
composition together according to Table~\ref{tab:endpoint-mix}. The
operating regimes should therefore be interpreted for this workload;
isolating endpoint count from load and class mix requires additional
controlled experiments.

\section{Conclusion}\label{sec:conclusion}

This paper presented a scalability study of 5G-TSN bridges under
heterogeneous industrial traffic, evaluating four MAC schedulers across
up to 40 endpoints, two bandwidths, and ideal and Jakes fading channels.
The results show three operating regimes separated by a saturation
threshold: below saturation, non-DRR schedulers behave similarly; near
saturation, QoS-aware scheduling sharply reduces critical-flow latency;
and under overload, QoS-PF is the only evaluated scheduler that
maintains near-complete delivery for the highest-priority traffic class.
Across the two evaluated bandwidths, the saturation threshold
approximately doubles when bandwidth doubles.

The study also shows that IEEE~802.1AS/gPTP traffic benefits from a
dedicated high-priority radio bearer, which prevents the clock-servo
divergence observed when synchronisation traffic shares the default
bearer. However, bearer isolation alone does not bound gPTP residence
time under saturation: endpoints carrying lower-priority data receive
fewer MAC scheduling opportunities, coupling synchronisation quality to
data-plane priority. Finally, the relative scheduler ranking is
preserved under the tested Jakes fading configuration, where fading is
associated with reduced critical-flow latency under saturation,
consistent with multi-user diversity.

For the evaluated configuration, these findings support QoS-aware MAC
scheduling, dedicated gPTP bearer provisioning, capacity planning below
the saturation threshold, and explicit consideration of the radio latency
floor when setting TSN deadlines. Future work will extend the analysis
to multi-cell deployments, mobility, additional channel models, higher
numerology, and configured-grant or semi-persistent scheduling for gPTP
protection under saturation.

\end{document}